# Behavioral and Performance Indicators of Depression and Anxiety in Electronic Learning Systems

Arya VarastehNezhad, Fattaneh Taghiyareh, *Department of Computer Engineering, University of Tehran*

***Abstract*—This study investigates whether behavioral and performance indicators derived from a Moodle-based learning management system are associated with university students' depression and anxiety in two undergraduate Computer Engineering courses. Using a quantitative observational design, LMS event logs, academic records, and self-reported Beck Depression Inventory-II and Beck Anxiety Inventory scores from 97 students were integrated. A broad set of behavioral and performance indicators spanning temporal engagement, session structure, deadline-related behavior, page-refresh patterns, and LMS navigation was extracted from raw event logs and analyzed using descriptive statistics, independent-samples t-tests with Benjamini-Hochberg FDR correction, effect sizes, and Spearman correlations; inventory scores were confirmed invariant by sex and academic year. Several indicators were significantly associated with depression and anxiety. Higher depression was associated with shifted temporal activity patterns, longer session durations, and shorter homework submission lead times, while higher anxiety was associated with concentrated temporal engagement and session-based differences. These findings suggest that routine LMS data can provide meaningful behavioral signals related to student well-being and may support earlier educational awareness of students who experience mental-health-related strain. At the same time, such indicators should be interpreted as contextual and non-diagnostic markers rather than as substitutes for clinical assessment.**

## I. INTRODUCTION

ELECTRONIC learning systems have become a central component of higher education, mediating a substantial portion of students' access to learning resources, assessment activities, and communication with instructors and peers. Learning management systems (LMSs) such as Moodle generate rich, timestamped interaction logs that record how, when, and how often learners engage with course content, assignments, quizzes, and discussion spaces. This data-intensive environment has created new opportunities for educational researchers to use behavioral traces not only to model academic engagement, but also to explore how digital learning behavior may reflect broader aspects of students' experiences in technology-mediated settings [1].

At the same time, student mental health has become a major concern in higher education, with depression and anxiety repeatedly identified as prevalent and impactful conditions among university students [2], [3]. Empirical studies have shown that depression and anxiety are associated with difficulties in motivation, concentration, time management, and academic persistence, which can in turn affect performance and progression [4]. In higher-education settings, depression symptoms may translate into irregular study routines, task avoidance, last-minute work, or prolonged but inefficient study episodes [5], [6], [7].

Because LMS-mediated learning generates detailed behavioral records, it is plausible that some of the ways in which depression and anxiety shape students' academic lives could be reflected in their digital activity patterns. Previous work in learning analytics has shown that LMS logs can be used to predict academic risk, identify students at risk of underperformance, and provide data-driven feedback through dashboards and early warning systems [8], [9], [10]. Related research outside formal e-learning contexts has further demonstrated that digital traces from games, wearable devices, smartphones, and social media can carry informative signals related to mental-health-related states [11], [12], [13], [14]. However, behavioral traces are at best indirect indicators and must not be interpreted as clinical diagnoses; rather, their potential value lies in supporting earlier awareness and more responsive educational support when used carefully and ethically [15], [16].

Despite this promise, there is no empirical work that quantitatively integrates validated depression and anxiety measures with detailed LMS behavioral indicators and academic-performance-related features within a single framework. Many studies on student mental health rely primarily on self-report instruments or interviews without incorporating fine-grained learning-behavior data, while many studies in learning analytics focus on digital traces and academic outcomes without grounding them in established mental-health instruments such as the Beck Depression Inventory-II (BDI-II) and the Beck Anxiety Inventory (BAI). This separation limits our ability to understand whether and how specific learner behaviors in LMS environments are meaningfully associated with depression and anxiety levels, and how these psychological variables relate to academic performance. A more integrated perspective is needed to examine mental-health measures, behavioral patterns, and performance outcomes together, particularly in authentic course settings where early educational awareness might inform supportive interventions rather than clinical judgments.

This study addresses this gap by examining how LMS-derived behavioral and performance indicators relate to depression and anxiety among undergraduate students in two LMS-mediated courses. Using questionnaire data from the BDI-II and BAI implemented within the LMS, together with detailed Moodle event logs and course performance records, the study investigates four research questions:

**RQ1.** What are the levels and distributions of depression and anxiety among students as measured by the BDI-II and BAI?

**RQ2.** Which LMS behavioral indicators (e.g., temporal activity patterns, session characteristics, deadline-related behavior) are significantly associated with students' depression levels?

**RQ3.** Which LMS behavioral indicators are significantly associated with students' anxiety levels?

**RQ4.** How do depression and anxiety levels affect students' academic-performance outcomes in the courses?

To address these questions, the study constructs a broad set of behavioral indicators from LMS logs, including overall activity volume, temporal engagement patterns, session-level characteristics, deadline-related behavior, and interaction breadth across course modules, and examines their relationships with depression and anxiety scores and course outcomes. The empirical findings indicate that several LMS-based indicators are significantly associated with students' depression and anxiety levels. In particular, shifted temporal activity patterns is associated with higher levels of depression, greater weekend activity is associated with higher levels of anxiety, and longer maximum session durations are related to higher depression. These results suggest that when students study, how they structure their LMS sessions, and how they distribute their activity across the week may provide informative signals about potential mental-health-related strain in e-learning environments, while remaining clearly distinct from clinical diagnosis.

This study contributes to the literature in three main ways. First, it extends learning analytics research beyond traditional academic prediction tasks by focusing explicitly on depression and anxiety as important outcomes and covariates in higher education, operationalized through validated self-report instruments. Second, it develops a multidimensional view of LMS behavior by integrating temporal, session-based, deadline-related, and performance-related indicators within a single quantitative framework grounded in authentic course data. Third, it provides empirical evidence that routine data generated in LMS environments can support early educational awareness of potential mental-health-related difficulties, thereby informing the design of more responsive and supportive learning-analytics approaches, while maintaining a clear conceptual boundary between behavioral indication and clinical assessment. The remainder of the paper is organized as follows. The next section reviews related work on student mental health, electronic learning systems, learning analytics in LMS environments, and behavior-based detection of psychological states through digital traces. The methodology section then describes the research design, study context, instruments, data sources, feature-engineering process, and analytical procedures. Subsequent sections present the results, discuss their implications and limitations, and outline directions for future research.

## II. Literature Review and Related Work

Research related to this study spans four areas: student mental health, electronic learning systems, learning analytics in LMS environments, and behavior-based detection of psychological states through digital traces. Depression and anxiety are among the most widely discussed mental-health concerns in student populations, with large-scale surveys and systematic reviews showing that these conditions are both prevalent and consequential in higher education [2], [3], [17]. Prior work has shown that depression and anxiety can affect emotional well-being, academic functioning, persistence, and help-seeking behaviors, highlighting them as major challenges in university contexts rather than marginal issues [18]. In addition, these conditions are influenced by broader contextual and social factors including socioeconomic status, stigma, workload, and work-life balance that shape whether psychological distress becomes visible, reported, and supported within educational institutions [19], [20], [21], [22].

The clinical literature emphasizes that depression is multidimensional, encompassing emotional, cognitive, behavioral, and physical symptoms such as low mood, hopelessness, fatigue, reduced interest, and cognitive slowing.

Anxiety-related difficulties, in turn, have been described in terms of excessive worry, physiological arousal, avoidance, impaired attention under threat, and altered information processing [23]. In educational settings, these symptom profiles may manifest as problems with concentration, motivation, energy regulation, sleep patterns, time management, and task completion, all of which are relevant to students' day-to-day academic routines. Empirical work has reported associations between depressive or anxious symptomatology and lower academic achievement, increased fatigue, and difficulties sustaining effort, suggesting that mental-health status may have both direct and indirect impacts on educational outcomes (positive or negative) [24], [25], [26].

Parallel to these developments, electronic learning systems and LMS platforms have become central infrastructures in higher education, evolving from simple content repositories into multifaceted environments that support content delivery, communication, assessment, feedback, personalization, and detailed activity tracking [27]. Widespread adoption of systems such as Moodle has been further accelerated by large-scale shifts toward blended and online learning, for example during the COVID-19 pandemic, which has reinforced the role of LMS platforms as primary mediators of students' learning experiences [1], [28]. Because LMSs generate structured, timestamped, and context-rich logs of navigation, content access, submissions, feedback viewing, and other course interactions, they provide an unusually detailed record of how students move through their courses over time [29], [30]. This has made LMS-based ecosystems a natural setting for educational data mining and learning analytics (LA).

A substantial body of LA research has used LMS data to model engagement, predict academic performance, and design early warning systems. Studies have shown that features such as login frequency, number of resource views, assignment activity, and patterns of participation can serve as predictors of course outcomes or dropout risk, and can support interventions aimed at improving success and retention [9], [31], [32]. Other work has focused on feedback and dashboard systems that translate LMS interaction data into actionable information for students and instructors, aligning with principles of formative assessment and self-regulated learning. Additional strands of research have examined LMS acceptance and continued-usage intentions, as well as adaptive and personalized learning pathways, open social student modelling, and personalized feedback systems [33], [34], [35]. Taken together, this literature demonstrates that LMS log data can be exploited to understand and support academic engagement, but it has predominantly focused on learning outcomes such as grades, retention, and participation rather than on students' mental health [36].

Studies of students' perceptions indicate both interest in and concern about the use of learning analytics for mental-health support, emphasizing issues such as privacy, consent, interpretability, and the risk of misclassification [16]. However, much of this emerging literature remains at a conceptual, qualitative, or small-scale level, and no studies implement validated clinical instruments (such as the BDI-II or BAI) in conjunction with detailed LMS behavioral indicators in authentic course settings.

Beyond educational contexts, a growing body of research has investigated the use of digital behavioral traces for mental-health detection and monitoring. Work in this domain has analyzed gameplay logs, smartphone and wearable-device data, smartwatch interactions, speech features, and social-media activity to identify patterns associated with depression, anxiety, or related conditions [11], [12], [13], [14], [37]. Systematic reviews of machine-learning approaches in digital mental-health research highlight both the promise and the challenges of using behavioral data to infer psychological states, noting issues such as data heterogeneity, model generalizability, and the need for rigorous validation [38]. These studies collectively suggest that human behavior in digital environments often contains signals that correlate with affective or psychological states, even when the technologies involved were not originally designed as mental-health tools.

At the same time, behavior-based mental-health detection raises important conceptual, ethical, and practical concerns. Behavioral traces are inherently indirect indicators and can be influenced by context, individual differences, measurement limitations, and social factors. Scholars have cautioned that analytics systems must avoid over-interpreting behavioral signals as diagnostic of mental disorders and should instead focus on supporting earlier awareness, more informed educational understanding, and ethically grounded interventions. In educational environments, this implies that LMS-based indicators should not be used for automated diagnosis, but can be explored as complementary markers that may help educators recognize patterns of engagement or strain that warrant attention, provided that appropriate safeguards, transparency, and support structures are in place [39].

Existing work at the intersection of learning analytics and wellbeing tends to focus either on conceptual frameworks, perceptions, or early prototypes rather than on systematically examining how specific LMS indicators relate to levels of depression and anxiety in real courses. To our knowledge, no studies have quantitatively combined detailed LMS behavioral indicators, validated depression and anxiety measures, and academic-performance data in authentic higher-education settings [38]. The present study contributes to this emerging area by bringing together these strands: it uses BDI-II and BAI scores collected through Moodle, a broad feature space of LMS-derived behavioral and performance indicators, and course outcomes to examine how depression and anxiety relate to e-learning behavior and academic performance, thereby addressing a gap at the intersection of learning analytics, behavior modelling, and student mental-health research.

## III. Methodology

### *A. Research Design*

This study employed a quantitative, data-driven observational design to examine whether behavioral traces recorded in an LMS-based e-learning environment were associated with students' depression and anxiety levels and with their academic-performance outcomes. The methodological workflow followed an input-process-output structure (Figure *1*) in which LMS event logs, questionnaire responses, academic records, and demographic information were integrated and transformed into a student-level analytical dataset of behavioral and performance indicators. The design was intended to address the four research questions stated in the Introduction by examining the distributions of depression and anxiety, identifying LMS-based indicators associated with these constructs, and investigating how depression and anxiety related to academic and performance-related outcomes.

The study was conducted in an educational rather than clinical context, and its aim was to identify interpretable behavioral and performance-related markers associated with depression and anxiety rather than to provide diagnosis or replace formal clinical assessment. Accordingly, the analytical strategy was exploratory and combined three complementary stages: descriptive statistical analysis, group-comparison analysis, and correlation analysis. Descriptive analysis was used to summarize questionnaire distributions and participant characteristics, while inferential analysis examined both continuous symptom scores and grouped symptom-status variables derived from the BDI-II and BAI.

To support this two-level analysis, depression and anxiety were represented in two forms throughout the study: as continuous questionnaire scores and as categorized status variables based on symptom-severity groupings. This made it possible to examine associations between LMS indicators and mental-health measures from two complementary perspectives: monotonic relationships with symptom severity and statistical differences between lower-symptom and higher-symptom student groups. Because the two course groups differed in size, structure, and grading composition, all analyses were conducted separately for each group so that group-specific behavioral patterns would not be obscured by pooling the datasets. This design allowed the study to move beyond simple description of LMS activity and toward a multidimensional examination of how temporal behavior, session characteristics, deadline-related actions, and performance-related LMS interactions were related to depression and anxiety in authentic course settings.

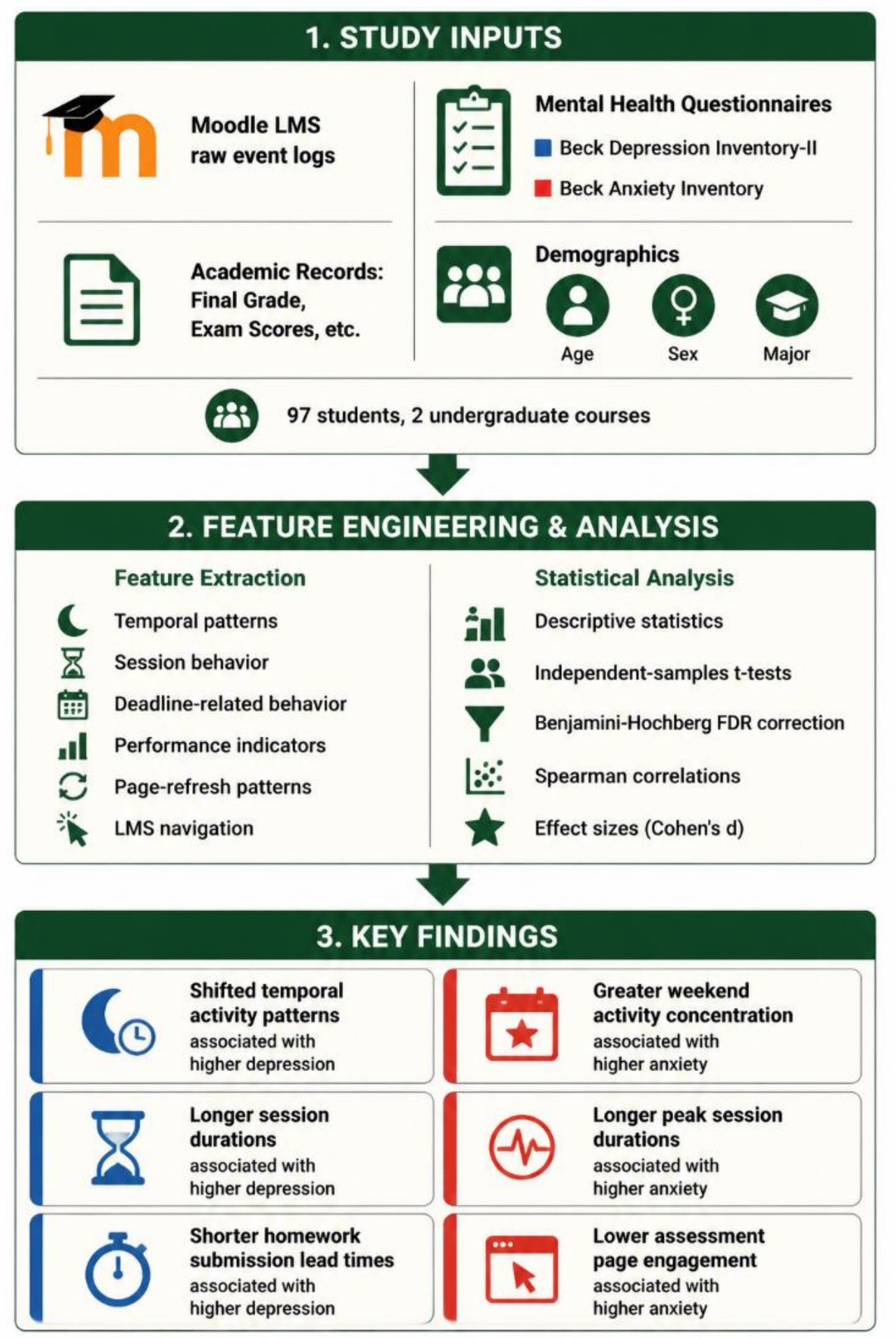

Figure 1: Methodological workflow of the study, illustrating the input-process-output structure in which LMS event logs, questionnaire responses, academic records, and demographic data were integrated and transformed into a student-level analytical data.

## *B. Study Context and Participants*

The study was conducted at a large public university in Iran and focused on two undergraduate Computer Engineering course groups supported through a Moodle-based blended learning environment. In both courses, Moodle was used to distribute learning materials, deliver online quizzes, deliver and collect assignment and project submissions, publish grades, and communicate course announcements, while regular face-to-face class meetings and on-site examinations were also maintained.

The initial sample included 104 enrolled students across the two course groups. Because the inferential analyses required complete mental-health data, only students who completed both the BDI-II and the BAI were retained in the final analytical dataset. After applying this criterion, the final sample used for analysis consisted of 60 students in Group 1 and 37 students in Group 2.

Participation in the questionnaires was voluntary and was encouraged through a small course incentive of 0.5 points out of 20, which was the same for all students who chose to participate. Questionnaire completion rates were high in both groups, reaching 90.90% in Group 1 and 97.36% in Group 2 for each instrument. To reduce the potential effect of the time of questionnaire administration on reported symptom levels, the questionnaires were distributed in two waves within each group: the first wave was administered between weeks 4 and 5 of the semester, and the second wave between weeks 8 and 9. Students were randomly assigned to one of the two waves. No statistically significant difference was found between the BDI-II and BAI scores of students in the two waves, supporting the assumption that administration timing did not systematically bias the self-reported symptom levels. LMS log data were collected across the full semester for all students regardless of their assigned questionnaire wave.

The two course groups also differed in academic seniority and course structure, which is relevant for interpreting behavioral and performance differences across groups. In the larger group, most students were in their fourth year of study, whereas in the smaller group, the majority were third-year students (Figure 2).

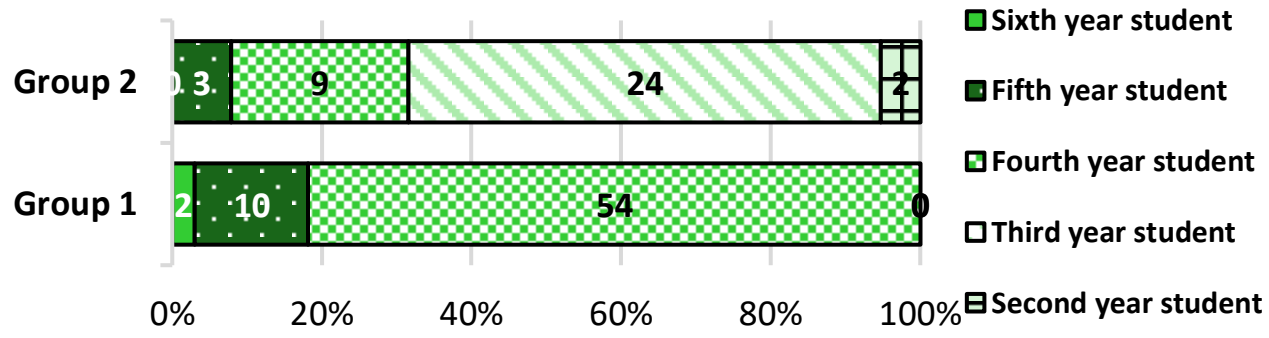

Figure 2: Distribution of students by academic year in Group 1 and Group 2.

At the beginning of each questionnaire, students were shown an information statement explaining that participation was voluntary, that the instrument was being used for research purposes only, that it was not intended for diagnosis or treatment, and that responses would be handled confidentially and analyzed in aggregated form (Figure 3).

This questionnaire consists of 21 groups of statements. Please read each group carefully and select the one statement that best describes how you have been feeling during the past two weeks, including today. If you believe that more than one statement within a group applies equally well to you, select the statement with the higher number. Please choose only one statement from each group. This instruction applies to all items, including Item 16 (Changes in Sleeping Pattern) and Item 18 (Changes in Appetite).

Estimated completion time: less than 5 minutes.

This questionnaire is administered solely for research purposes. It is not intended to provide a diagnosis, clinical assessment, or treatment recommendation.

This questionnaire also consists of 21 groups of statements. Please read each group carefully and select the one statement that best describes your experience during the past two weeks, including today. If you believe that more than one statement within a group applies equally well to you, select the statement indicating the greater severity.

Estimated completion time: less than 4 minutes.

This questionnaire is administered solely for research purposes. It is not intended to provide a diagnosis, clinical assessment, or treatment recommendation.

Figure 3: Information statement displayed to students at the beginning of each questionnaire, clarifying voluntary participation, research purpose, and confidentiality of responses.

*C. Data Sources and Collection Procedure*

Four main data sources were integrated at the student level in this study: Moodle event-log data, responses to the BDI-II and BAI questionnaires, academic-performance records, and demographic information. These sources were linked through anonymized student identifiers so that behavioral, psychological, academic, and background variables could be analyzed within a single student-level framework.

The first data source consisted of raw event logs automatically generated by students' interactions with the Moodle learning environment during the semester. The exported log files contained event-level operational fields such as timestamp, username, affected user, course module, activity component, event type, descriptive text, and event source, with each row representing one recorded LMS event. These low-level records formed the basis for deriving higher-level behavioral indicators such as activity volume, temporal engagement, session structure, deadline-related behavior, interaction patterns, and page-refresh behavior.

The second data source comprised self-report mental-health measures collected through Moodle-based implementations of the Beck Depression Inventory-II (BDI-II) [40] and the Beck Anxiety Inventory (BAI) [41]. Each questionnaire was administered within the LMS through an instruction page followed by structured item pages, and the resulting scores were retained both as continuous variables and as categorized symptom-severity variables for later analysis. BDI II scores were interpreted in four standard bands: minimal depression (0-13), mild depression (14-19), moderate depression (20-28), and severe depression (29-63) [42], while BAI scores were interpreted as minimal anxiety (0-7), mild anxiety (8-15), moderate anxiety (16-25), and severe anxiety (26-63). In addition, for inferential analyses that compared lower-symptom and higher-symptom students, these four-level scales were collapsed into two broader categories: minimal/mild versus moderate/severe.

The third data source consisted of academic-performance records for both course groups. These records included final grades as well as detailed component scores such as examinations, quizzes, assignments, projects, and, where applicable, laboratory-related performance. The fourth data source comprised demographic information retained in limited form for descriptive and analytical purposes. These variables included gender, academic year, and whether Computer Engineering was the student's primary field of study. Across both groups combined, the sample comprised 97 students, of whom 78 were male (80.4%) and 18 were female (18.6%).

*D. Behavioral and Performance Indicator Extraction*

After data collection and initial cleaning, the raw Moodle event-log data were transformed into a secondary student-level dataset suitable for quantitative analysis. While the exported logs contained only nine operational fields per event, the feature-engineering process derived a broad set of interpretable indicators that summarized how each student interacted with the LMS across multiple behavioral dimensions. Each row in the resulting dataset corresponded to a single student in a single course group, and each column represented one extracted feature.

A first group of indicators captured the overall volume and continuity of student activity. These included the total number of events generated by each student (a basic measure of engagement), the number of unique active days (days on which at least one event was recorded), and the number of distinct course modules visited at least once. For each of these quantities, a class-relative rank was also computed (e.g., the rank of total logs within the class), so that each feature was represented both in absolute form and as a within-course position measure. The use of ranks was motivated by the fact that absolute values for some features can be influenced by course-specific settings or structural differences (such as the number of activities or grading components), whereas within-course ranks are less sensitive to such differences and emphasize how a student's behavior compares to peers in the same context.

Temporal engagement patterns were summarized by standardizing date and time values and categorizing event timestamps into four daily time periods: morning (06:00-12:00), afternoon (12:00-18:00), evening (18:00-00:00), and night (00:00-06:00). For each student, the number of events in each period was counted; these counts were also normalized by dividing by the student's total number of logs to obtain relative frequencies. Class-relative ranks for each time-of-day variable were computed to capture each student's position within the course on those indicators. Additional features measured the intensity of activity on active days (mean number of events on such days) and on the busiest day of the semester for each student.

A further group of indicators focused on assignment and deadline-related behavior. For activities that had submission deadlines, the lead time before submission was calculated as the time difference between each student's submission timestamp and the corresponding deadline, expressed in hours or minutes. Let $t_{ij}^{sub}$ and $t_j^{dead}$ denote the submission time of student *i* and the deadline time for activity *j*, respectively. The raw lead time was

$$\Delta t_{ij} = t_j^{\text{dead}} - t_{ij}^{\text{sub}}.$$

Because raw lead times can span a wide range, from several weeks before the deadline to hours after it, the distribution is typically heavy-tailed and negatively skewed. A small number of students who submitted very early exert disproportionate influence on group means and inflate variance, potentially masking more meaningful behavioral differences among students who submitted close to the deadline. To address this, a sign-preserving logarithmic transformation was applied to each raw lead time value. This transformation compresses extreme values symmetrically on both sides of the deadline while preserving the direction of the lead time. For each student, aggregate indicators such as the mean raw lead time and the mean transformed lead time across all relevant activities were then computed. Similar features were extracted for the delay between the posting of a new activity and the student's first view of that activity.

Deadline-related concentration was examined by calculating, for each student, the number of events generated on deadline days, the ratio of deadline-day events to total events, and the class-relative rank of that ratio. Weekly temporal behavior was also summarized by separating events on working days from events on weekends and computing the ratio of weekend events to total events, together with the corresponding class rank.

Session-based indicators were derived from the event logs by identifying contiguous periods of activity separated by extended inactivity. For each student *i*, events were ordered by timestamp $t_{i1}, t_{i2}, \ldots, t_{in_i}$ and the time differences

$$\Delta t_{ik} = t_{ik} - t_{i,k-1} \quad (k = 2, \ldots, n_i)$$

were computed. A new session was assumed to start whenever $\Delta t_{ik}$ exceeded eight hours or when $k = 1$; all events between two such break points were grouped into a single session. Formally, the session indicator $s_{ik}$ was defined as

$$s_{ik} = \begin{cases} 1, & k = 1 \, or \, \Delta t_{ik} > 8 \, hours, \\ 0, & otherwise. \end{cases}$$

Each contiguous sequence of events with $s_{ik} = 0$ following a starting point with $s_{ik} = 1$ formed a session for student *i*. For each detected session *m*, the start time $t_{im}^{\text{start}}$, end time $t_{im}^{\text{end}}$, and duration

$$d_{im} = t_{im}^{\text{end}} - t_{im}^{\text{start}}$$

were calculated. These session records were then aggregated at the student level to produce features such as the total number of sessions $N_i^{\text{sess}}$, the total time spent in the LMS across all sessions $D_i^{\text{tot}} = \sum_m d_{im}$, the mean session duration, and the maximum session duration. Class-relative ranks were computed for $N_i^{\text{sess}}$, $D_i^{\text{tot}}$, the average session duration, and the maximum session duration. In the learning-analytics literature, inactivity thresholds for defining sessions vary and are often chosen empirically based on the distribution of inactivity periods or course design. To ensure the chosen threshold was appropriate for the present data, five candidate thresholds were evaluated: 2, 4, 6, 8, and 10 hours. Thresholds of 2 and 4 hours produced a large number of consecutive short sessions for many students, suggesting that normal within-study pauses were being incorrectly fragmented into separate sessions. Conversely, thresholds of 10 hours caused the session count to drop substantially for a significant portion of students, collapsing extended multi-day engagement patterns into single sessions and losing temporal granularity. The 6-hour threshold produced results comparable to 8 hours for most students, but the 8-hour threshold yielded more stable and consistent session counts across both groups and aligned more naturally with day-scale study patterns.

Additional behavioral markers related to login and page-refresh behavior were also extracted. New login events were identified by examining time gaps between successive events for each student and marking intervals longer than eight hours as indicating a new login. If $\Delta t_{ik} > 8 \, hours$, then a new-login indicator $\ell_{ik} = 1$; otherwise $\ell_{ik} = 0$, and the total number of new logins for student *i* was computed as $L_i = \sum_k \ell_{ik}$. Page refreshes were detected by grouping events by student and event description, again ordering them by timestamp, and computing within-group time differences. For events with the same description and component, a refresh indicator $r_{ik}$ was defined as

$$r_{ik} = \begin{cases} 1, & \Delta t_{ik} < 20 \, seconds, \\ 0, & otherwise. \end{cases}$$

and the total number of detected refreshes for student *i* was then $R_i = \sum_k r_{ik}$. In existing LMS and EDM research, session and login-based indicators are common, whereas explicit identification of page refreshes is rarely reported; thus, the refresh heuristic used here represents a pragmatic extension tailored to the present dataset rather than an established standard.

Alongside behavioral indicators, performance-related variables were incorporated into the feature set so that psychological measures could be examined in relation to educational outcomes. These included final course grades and selected component scores (e.g., examinations, assignments, projects, laboratory grades), as well as relative ranks for some of these measures within each course group.

*E. Statistical Analysis*

The analytical strategy was designed to address the four research questions by combining descriptive statistics, group-comparison analyses, and non-parametric correlation analyses. For RQ1, descriptive statistics and visualizations were used to summarize the distributions of BDI-II and BAI scores, both as continuous variables and grouped into the four standard severity bands and the two aggregated categories (minimal/mild vs. moderate/severe). For RQ2 and RQ3, the analysis focused on identifying associations between depression and anxiety measures and the extracted LMS behavioral indicators, while RQ4 examined how depression and anxiety levels were related to academic-performance outcomes and performance-related LMS behavior.

For the grouped comparisons, independent-samples t-tests were used to examine whether the mean values of selected indicators differed significantly between the two groups within each course. The independent-samples t-test evaluates whether the observed difference between two group means is large relative to the variability within the groups. Because statistical significance alone does not indicate the magnitude of a difference, effect sizes were also reported for the grouped analyses. Cohen's *d* was used to quantify the standardized mean difference between two groups and can be expressed as

$$d = \frac{\bar{x}_1 - \bar{x}_2}{s_p},$$

where $s_p$ is the pooled standard deviation of the two groups. Cohen's *d* provides an estimate of how large the group difference is in standard-deviation units, making it easier to interpret the practical importance of statistically significant and non-significant findings alike [43]. Reporting both p-values and effect sizes therefore allowed the analysis to consider not only whether a difference was statistically detectable, but also whether it was substantively meaningful.

Given that some behavioral indicators exhibited skewed distributions and that BDI-II and BAI scores can be treated as at least ordinal with non-normal distributions, Spearman's rank correlation coefficient ($\rho$) was used as the main association measure. Spearman's $\rho$ assesses the strength and direction of monotonic relationships between two variables by correlating their rank orders rather than their raw values, making it robust to non-linearities and outliers [44]. When there are no tied ranks, and $d_i$ denotes the difference between the ranks of the two variables for observation $i$ (for $n$ observations), the coefficient can be expressed as

$$\rho = 1 - \frac{6\sum_{i=1}^{n} d_i^2}{n(n^2-1)}.$$

More generally, Spearman's $\rho$ can be obtained by computing the Pearson correlation coefficient on the ranked values of the two variables. For each pair of variables of interest, such as a behavioral feature and a depression or anxiety score both the raw feature values and their class-relative ranks were considered, reflecting the dual representation of behavior in absolute and within-course positional terms. Depressive and anxious symptom levels, academic-performance measures, and demographic variables were included in the correlation analysis.

All inferential analyses were conducted separately for the two course groups in order to preserve group-specific behavioral and performance patterns and to avoid obscuring differences arising from course structure or grading composition. A conventional significance threshold of $\alpha$ = 0.05 (two-tailed) was used to identify statistically significant results. To control the expected proportion of false discoveries among the significant results, the Benjamini-Hochberg (BH) false discovery rate (FDR) procedure was applied to all t-test p-values [45]. Under this procedure, p-values are ranked from smallest to largest (rank k out of m total tests), and each is compared against the adaptive threshold $\frac{k}{m} \times \alpha$, where $\alpha$=0.05. The BH-adjusted p-value for each finding is $\tilde{p}_k = min_{j\geq k}\left(\frac{m}{j} \cdot p_j\right)$, enforcing monotonicity. With m = 25 t-tests, all 25 findings retained significance after FDR correction (all $\tilde{p} < 0.05$), indicating that the reported associations are robust to multiple-comparison adjustment. All statistical analyses were performed in Python using standard scientific and statistical libraries. A consolidated summary of all indicators yielding significant or noteworthy results, including group means, percentage differences, p-values, Cohen's d, and Spearman correlations is provided in Appendix A for reference.

## IV. Results

### A. Questionnaire distributions and participant profile

The final analytical sample included 60 students in Group 1 and 37 students in Group 2 who completed the depression and anxiety questionnaires. The findings reported in this section are summarized in full in Appendix A. For each indicator, the table provides mean values for non-affected and affected students, the percentage difference, the t-test p-value, Cohen's d, and Spearman correlations with both the continuous BDI/BAI inventory scores and the binary clinical status label. The distribution of depression and anxiety severity levels across the two groups is presented in Figure 4.

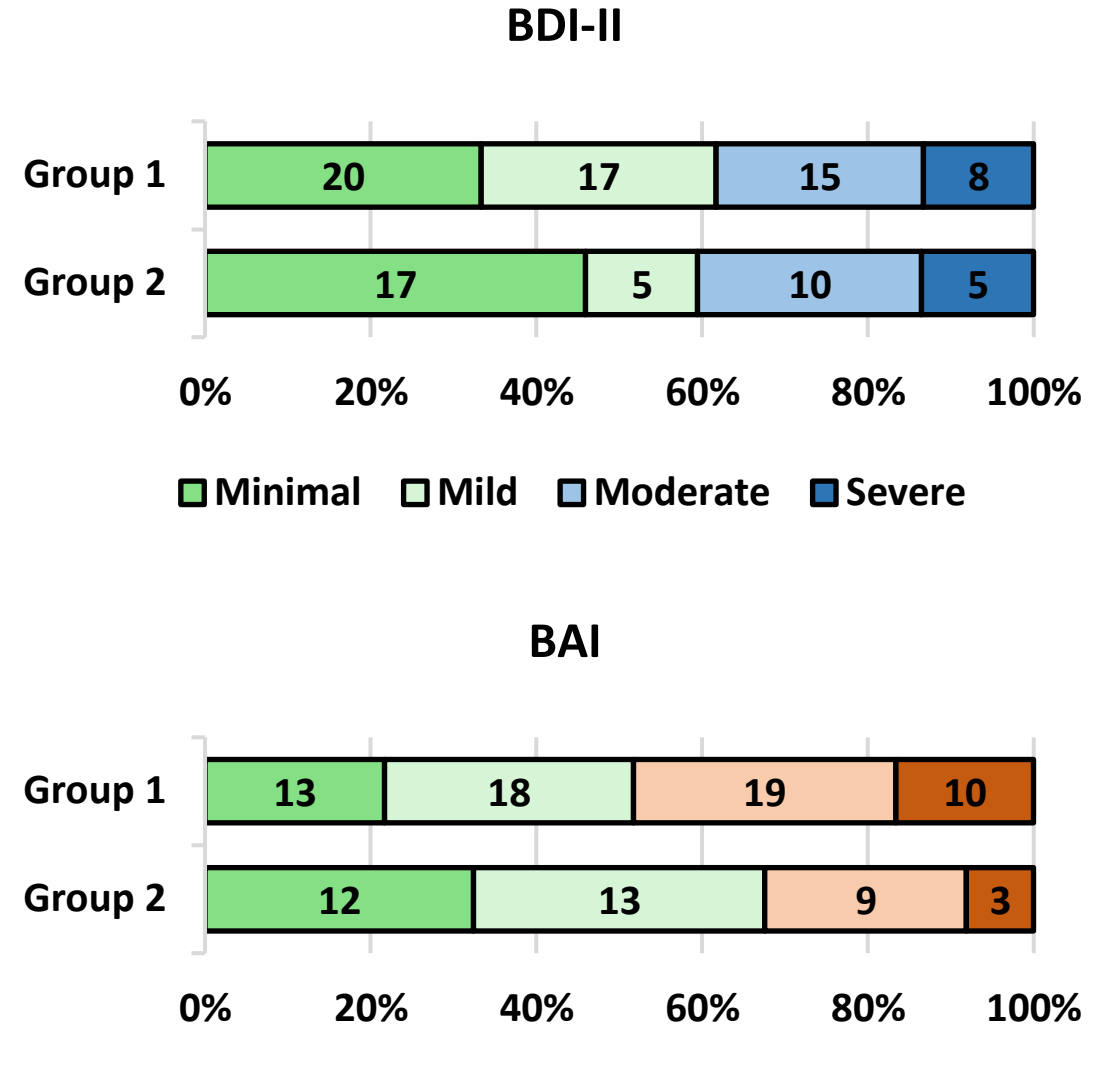

Figure 4: 100% stacked bar plots showing the distribution of students across minimal, mild, moderate, and severe categories for depression (BDI-II) and anxiety (BAI) in Group 1 and Group 2.

Depression and anxiety scores were positively related in both course groups (Figure 5). Spearman correlation analysis showed a positive association between BDI-II and BAI scores in Group 1 ($\rho$=0.493, $p$=0.000064, $n$ = 60) and Group 2 ($\rho$=0.538, $p$=0.000594, $n$ = 37). These results indicate that higher depression scores tended to co-occur with higher anxiety scores across both groups.

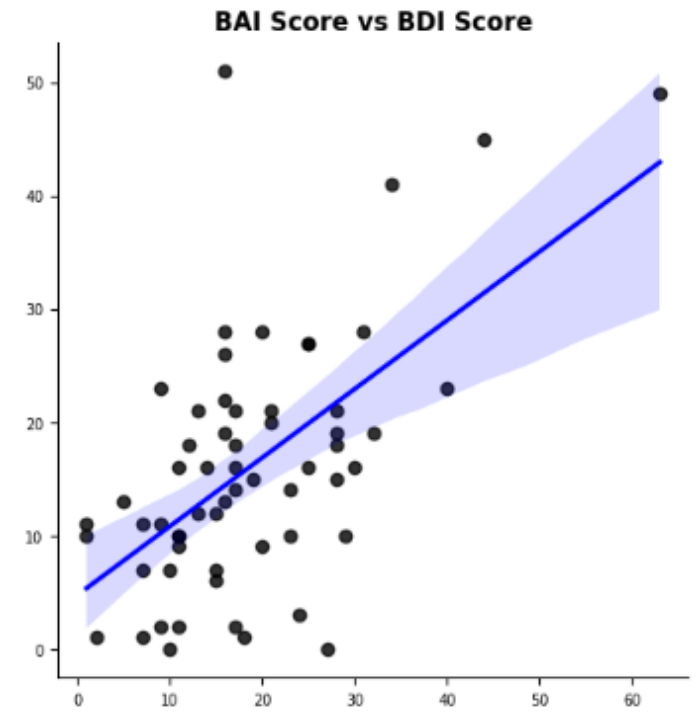

Figure 5: Scatter plot of BAI scores versus BDI-II scores, illustrating the positive association between depression and anxiety scores.

Neither BDI-II nor BAI scores differed significantly by sex in either group (Group 1 BDI-II: MWU $p$=0.393; BAI: MWU $p$=0.136; Group 2 BDI-II: MWU $p$=0.889; BAI: MWU $p$=0.316). Although female students in Group 1 scored descriptively higher on both scales, none of these differences reached significance, and the female subgroup in Group 2 comprised only three students. Accordingly, sex was not treated as a covariate in the subsequent analyses. BDI-II scores did not differ significantly by academic year in either group (Group 1: Kruskal-Wallis $H$=0.433, $p$=0.806; Group 2: $H$=5.547, $p$=0.136), nor did BAI scores in Group 1 ($H$=0.539, $p$=0.764). In Group 2, a statistically significant BAI difference was detected ($H$=8.929, $p$=0.030), but post-hoc analysis traced this entirely to the three Year 5 students ($M$=29.00), whose scores were significantly higher than Year 3 (MWU $p$=0.013) and Year 4 peers (MWU $p$=0.016); given $n$=3, this result is statistically fragile. Academic year was therefore not treated as a covariate in the behavioral analyses.

*B. Overall LMS activity indicators*

Among the overall LMS activity indicators, refresh-related behavior showed the clearest association with depression in Group 1. Depressed students had lower refresh counts than non-depressed students, with mean refresh counts of 41.83 (SD=21.41) versus 55.43 (SD=24.41), corresponding to a mean difference of -13.61 and a percentage difference of -24.55% ($p$=0.0320, $d$=-0.584). A similar pattern was observed for refresh-count ratio, which was also lower among depressed students than non-depressed students, with means of 9.03 (SD=3.28) versus 10.88 (SD=2.44), a mean difference of -1.85, and a percentage difference of -17.00% ($p$=0.0153, $d$=-0.664) (Figure 6). Spearman analysis was consistent with these group comparisons, showing negative associations between depression status and refresh count ($\rho$=-0.288, $p$=0.0256) and between depression status and refresh-count ratio ($\rho$=-0.268, $p$ =0.0383).

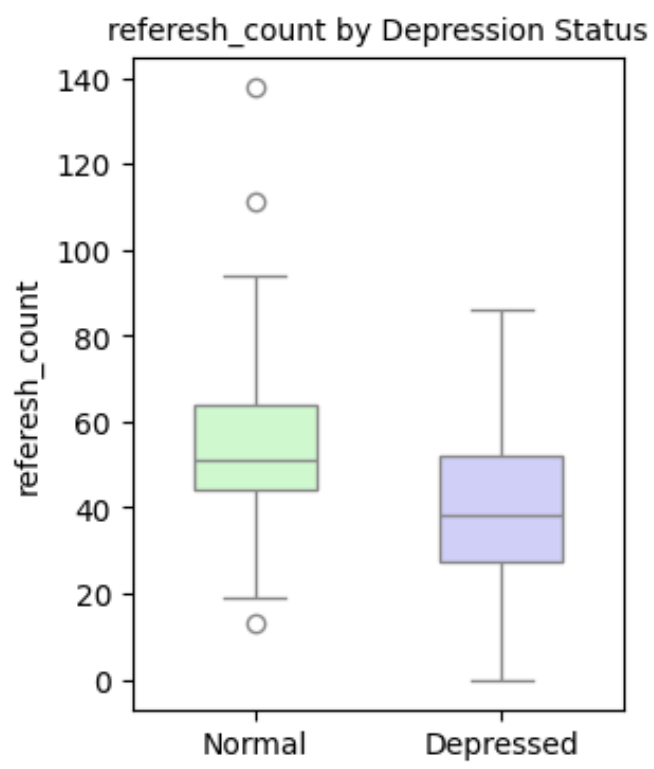

Figure 6: Group differences in LMS page-refresh behavior between depressed and non-depressed students in Group 1.

*C. Temporal engagement patterns*

Temporal engagement patterns showed different associations with depression and anxiety. In Group 1, anxiety was associated with greater concentration of LMS activity in the afternoon. Specifically, anxious students had higher afternoon log counts than non-anxious students, with mean values of 191.28 versus 143.87, corresponding to a mean difference of 47.41 and a percentage difference of 32.95% ($p$=0.0084, $d$=0.705). Anxious students also had a higher ratio of afternoon logs than non-anxious students, with means of 40.63 versus 32.17, a mean difference of 8.47, and a percentage difference of 26.32% ($p$=0.0128, $d$=0.663). Depression showed a similar pattern in Group 1, where depressed students had a higher ratio of afternoon logs than non-depressed students, with means of 42.37 versus 32.47 a mean difference of 9.90, and a percentage difference of 30.50% ($p$=0.0043, $d$=0.789). Correlation analysis supported these findings, showing positive associations between anxiety status and afternoon log count ($\rho$=0.343, $p$=0.0073), anxiety status and ratio of afternoon logs ($\rho$=0.321, $p$=0.0125), BAI score and ratio of afternoon logs ($\rho$=0.254, $p$=0.0499), depression status and ratio of afternoon logs ($\rho$=0.371, $p$=0.0035), and BDI-II score and ratio of afternoon logs ($\rho$=0.394, $p$=0.0018).

Additional temporal patterns in Group 1 suggested lower late-day activity among depressed students. Depressed students had lower evening log counts than non-depressed students, with mean values of 147.17 versus 201.32, corresponding to a mean difference of -54.15 and a percentage difference of -26.90% ($p$=0.0315, $d$=-0.585). Consistent with this pattern, depression status was negatively correlated with both night log count ($\rho$=-0.272, $p$=0.0353) and evening log count ($\rho$=-0.261, $p$=0.0437) in Group 1.

In Group 2, depression was associated with increased nighttime activity and greater weekend concentration. Depressed students had higher night log counts than non-depressed students, with mean values of 59.73 versus 26.23, corresponding to a mean difference of 33.51 and a percentage difference of 127.75% ($p$=0.0028, $d$=1.075). They also had a higher ratio of night logs, with means of 16.80 versus 7.32, a mean difference of 9.48, and a percentage difference of 129.57% ($p$=0.0167, $d$=0.842). Weekend-related indicators showed the same direction, as depressed students had higher weekend log counts than non-depressed students, with means of 190.13 (SD=88.28) versus 136.91, a mean difference of 53.22, and a percentage difference of 38.88% ($p$=0.0245, $d$=0.787), and a higher weekend log ratio, with means of 0.481 versus 0.395, a mean difference of 0.086, and a percentage difference of 21.86% ($p$=0.0283, $d$=0.766). Correlation analysis supported these findings, showing positive associations of BDI-II score with night log count ($\rho$=0.424, $p$=0.0090), ratio of night logs ($\rho$=0.409, $p$=0.0119), and weekend log ratio ($\rho$=0.371, $p$=0.0239), as well as positive associations of depression status with night log count ($\rho$=0.410, $p$=0.0117), ratio of night logs ($\rho$=0.380, $p$=0.0205), and weekend log ratio ($\rho$=0.343, $p$=0.0375).

Anxiety in Group 2 was most clearly associated with weekend activity concentration. Anxious students had a higher weekend log ratio than non-anxious students, with mean values of 0.491 versus 0.401, corresponding to a mean difference of 0.090 and a percentage difference of 22.46% ($p$=0.0292, $d$=0.798). This result was consistent with the correlation analysis, which showed a positive association between BAI score and weekend log ratio ($\rho$=0.501, $p$=0.0016) and between anxiety status and weekend log ratio ($\rho$=0.363, $p$=0.0273).

*D. Session-based engagement patterns*

Session-based indicators showed their strongest associations with depression in Group 2. Depressed students in this group had substantially longer maximum sessions than non-depressed students, with mean values of 67,099 versus 40,979, corresponding to a mean difference of 26,120 and a percentage difference of 63.74% ($p$=0.0029, $d$=1.072). This pattern was consistent with the correlation analysis, which showed a positive association between depression and maximum session duration both for BDI-II score ($\rho$=0.498, $p$=0.0017) and for depression status ($\rho$=0.495, $p$=0.0018). Depressed students in Group 2 also had greater total session duration than non-depressed students, with mean values of 227,377 versus 137,559, a mean difference of 89,817.51, and a percentage difference of 65.29% ($p$=0.0082, $d$=0.938). In addition, average session duration was higher among depressed students, with mean values of 7,461.87 versus 5,027.64, corresponding to a mean difference of 2,434.23 and a percentage difference of 48.42% ($p$=0.0099, $d$=0.914), and depression status was also positively correlated with average session duration ($\rho$=0.392, $p$=0.0165).

Anxiety in Group 2 was also associated with distinctive session-related patterns. Anxious students had a longer largest inactive period than non-anxious students, with mean values of 18.50 versus 12.56, corresponding to a mean difference of 5.94 and a percentage difference of 47.29% ($p$=0.0082, $d$=0.984). They also had longer maximum session durations, with mean values of 64,817.08 versus 45,209.04, a mean difference of 19,608.04, and a percentage difference of 43.37% ($p$=0.0391, $d$=0.753).

In Group 1, session-related effects were more limited, but anxiety was again associated with longer average sessions. Anxious students had a higher average session duration than non-anxious students, with mean values of 6,949.72 versus 5,529.07, corresponding to a mean difference of 1,420.66 and a percentage difference of 25.69% ($p$=0.0293, $d$=0.577). Overall, these results suggest that depression and anxiety were associated not simply with how often students used the LMS, but also with how their learning activity was organized into longer or more prolonged sessions.

### *E. Deadline-related behaviors*

Deadline-related indicators showed their clearest pattern in Group 1, where depression was associated with shorter submission lead times for practical homework activities. Across both practical homework assignments, depressed students submitted closer to the deadline than non-depressed students. For the first practical homework, the depressed group had a mean submission lead time of -3903 seconds, compared with 3272 seconds in the non-depressed group, corresponding to a mean difference of -3663 minutes ($p$=0.0031, $d$=-0.818), and depression score was also negatively correlated with this submission lead time ($\rho$=-0.284, $p$=0.0280). The same general pattern appeared for the second practical homework, where depressed students again showed a shorter submission lead time than non-depressed students, with mean values of 790 minutes versus 4603 minutes, a mean difference of -3812 minutes ($p$=0.0378, $d$=-0.565), and depression status was negatively correlated with this lead-time measure ($\rho$=-0.254, $p$=0.0498). A weighted, log-transformed version of the first practical-homework lead-time measure showed the same direction, with mean values of 1.55 for depressed students and 2.87 for non-depressed students ($p$=0.0474, $d$=-0.546), while depression score was again negatively correlated with this transformed indicator ($\rho$=-0.274, $p$=0.0355). Taken together, these results suggest a repeated pattern in which depressed students in Group 1 tended to submit practical homework closer to the deadline.

In Group 2, anxiety showed a weaker but still noteworthy relationship with homework-timing behavior. Anxiety status was positively correlated with the submission lead-time measure for one homework activity ($\rho$=0.390, $p$=0.0172), and the log-transformed version of the same indicator showed a similar positive association ($\rho$=0.385, $p$=0.0185). However, the corresponding group comparison for the transformed lead-time measure did not reach conventional statistical significance, with mean values of 2.733 (SD=0.197) for anxious students and 2.556 (SD=0.332) for non-anxious students ($p$=0.0970, $d$=0.599). This pattern may therefore be suggestive, but it should be interpreted cautiously.

Other deadline-related indicators were less reliable and should be treated as tentative. In Group 2, the average delay between homework posting and students' first view of the homework was higher among anxious students than non-anxious students, with mean values of 16161 minutes versus 10710 minutes, but this difference was not statistically significant ($p$=0.1279, $d$=0.548). Similarly, the number of LMS logs generated on deadline days was higher among depressed students than non-depressed students, with mean values of 107.07 versus 78.18, but this effect was only borderline ($p$=0.0542, $d$=0.667). The ratio of deadline-day logs to total logs showed the same direction, with mean values of 0.290 versus 0.230, but this difference was not statistically significant ($p$=0.1266, $d$=0.524). These latter findings may indicate potentially meaningful tendencies, but they are not robust enough to be treated as firm results.

### *F. Academic performance indicators*

Direct academic-outcome differences were limited in the present study, but several observations are still worth reporting. In Group 2, both anxiety and depression were associated with marginally higher project scores rather than lower ones. Anxious students had a mean project score of 98.42 compared with 81.44 for non-anxious students ($p$=0.0852, $d$=0.622), while depressed students had a mean project score of 97.07 compared with 80.05 for non-depressed students ($p$=0.0696, $d$=0.627). Although these differences did not reach conventional statistical significance, they indicate that elevated depression or anxiety did not necessarily coincide with poorer project performance in this course group.

Clearer performance-related patterns appeared in how students interacted with assessment and feedback components in the LMS. In Group 1, higher anxiety was associated with lower engagement with quiz-related pages, especially quiz-attempt-report pages, with negative correlations for both anxiety score ($\rho$=-0.458, $p$=0.00023) and anxiety status ($\rho$=-0.377, $p$=0.0030), and anxiety score was also negatively related to visits to the quiz page itself ($\rho$=- 0.258, $p$=0.0470). In the same group, higher depression was associated with lower engagement with performance-feedback pages, including negative correlations between depression score and visits to a homework grade-announcement page ($\rho$=-0.293, $p$=0.0231), between depression status and that same type of grade-announcement page ($\rho$=-0.333, $p$=0.0094), and between depression status and visits to the final-project page ($\rho$=-0.334, $p$=0.0091).

In Group 2, depression was associated with a different performance-related navigation pattern. Higher depression was linked to a lower proportion of exam-related activity, as shown by negative correlations for both depression score ($\rho$=-0.391, $p$=0.0166) and depression status ($\rho$=-0.361, $p$=0.0282). At the same time, higher depression was associated with more frequent visits to homework-solution and grade-related pages, with positive correlations ranging from about 0.33 to 0.39 and p-values between 0.0169 and 0.0474. More specifically, depression score was positively related to homework-solution-page visits ($\rho$=0.390, $p$=0.0169), homework-solution access more generally ($\rho$=0.360, $p$=0.0285), homework grade-announcement visits ($\rho$=0.357, $p$=0.0299), and final grade-announcement visits ($\rho$=0.333, $p$=0.0442), while depression status showed similar positive associations with homework-solution-related page visits ($\rho$=0.343, $p$=0.0375; $\rho$=0.342, $p$=0.0385; $\rho$=0.328, $p$= 0.0474). Overall, the clearer signals in this part of the analysis appeared in performance-related LMS behavior rather than in direct academic outcomes alone.

## V. Discussion

The present study showed that behavioral and performance-related indicators derived from an electronic learning system were meaningfully associated with students' depression and anxiety levels, but the clearest signals did not lie in simple differences in total activity or grades alone. Instead, the most informative patterns appeared in the timing of engagement, session structure, submission timing, and navigation of assessment-related LMS components. Although depression and anxiety were positively related, their behavioral correlates were not identical across the two course groups, suggesting that LMS-based signals of student vulnerability are meaningful but context-dependent. Overall, the findings support the view that routine LMS data may provide useful contextual information about student well-being, while also underscoring that such patterns are behavioral indicators rather than diagnostic evidence.

Across both groups, mental-health-related differences were reflected more clearly in how students engaged with the LMS than in whether they were simply active or inactive. Temporal and session-based patterns were especially informative: in Group 1, higher depression and anxiety were more closely associated with afternoon-concentrated activity, whereas in Group 2, depression was linked to higher nighttime and weekend activity as well as longer sessions, and anxiety was linked to stronger weekend concentration. These differences suggest that the relationship between mental health and study timing is unlikely to follow a single universal profile and should be interpreted in relation to course context, workload structure, and learner adaptation.

The session-related findings extend this interpretation by showing that psychological strain may influence not only when students study, but also how their engagement is organized across study episodes. In Group 2, higher depression was associated with longer average and maximum sessions and greater total session duration, while anxiety was associated with longer inactive gaps and longer peak sessions, suggesting more prolonged or less stable engagement rather than simple withdrawal from the LMS. Deadline-related behavior showed a similarly meaningful pattern: students with higher depression, especially in Group 1, tended to submit practical assignments closer to the deadline, whereas anxiety showed weaker and less stable associations with submission timing. Together, these findings indicate that time-of-day activity, session structure, and deadline proximity may reveal compressed or pressure-driven forms of engagement that remain difficult to detect through final submission status or overall participation counts alone.

Performance-related findings further showed that direct academic outcomes were less informative than performance-related LMS behavior. In Group 2, elevated depression and anxiety were not associated with clear academic decline and were even marginally related to higher project scores, but clearer differences appeared in how students used assessment-related resources. In Group 1, higher anxiety was associated with lower engagement with quiz-feedback pages, and higher depression was associated with lower engagement with grade-announcement and final-project pages; in Group 2, depression was associated with lower exam-related activity but greater engagement with homework-solution and grade-related pages. This suggests that academic success and healthy engagement should not be treated as equivalent, because students may maintain acceptable outcomes while relying on less stable, more compensatory, or more pressure-driven patterns of LMS use.

Taken together, the results suggest that the most informative LMS indicators of depression and anxiety were structural rather than purely volumetric: when students studied, how long they remained active, how close they worked to deadlines, and how they navigated evaluative course resources. The main contribution of the study therefore lies in showing that digital learning behavior may contain useful signals of student well-being that complement, rather than replace, conventional academic indicators. At the same time, the differences across the two groups show that these signals should be treated as contextual and probabilistic rather than as universal markers of depression or anxiety.

The findings of this study have both conceptual and practical implications for research and practice in learning analytics and student support. Conceptually, they suggest that LMS-based engagement should not be interpreted only in terms of academic productivity, because students with higher depression or anxiety may remain active and even academically successful while still showing less healthy temporal and behavioral patterns. In particular, night-heavy activity, weekend-shifted engagement, prolonged sessions, and deadline-proximal submissions appear to reflect forms of strain that are not captured well by grades or total log counts alone. This supports a distinction between achievement-related engagement and well-being-related engagement in digital learning environments.

The practical implication is not that LMS data should be used to diagnose students, but that it may help support earlier and more context-sensitive awareness. A combination of behavioral patterns such as increased night-time work, unusually long sessions, stronger weekend concentration, and repeated last-minute submissions could be used to trigger low-stakes check-ins, supportive outreach, or invitations to use available resources rather than automated labeling or punitive action. For instructors and academic advisors, dashboards that visualize temporal behavior alongside performance data may be more informative than systems that focus only on grades or participation volume. At the same time, any such implementation would require explicit consent, strong privacy safeguards, transparent governance, and close collaboration with mental-health professionals so that behavioral indicators are not overinterpreted or stigmatizing.

Several limitations should be considered when interpreting the results. First, the sample was drawn from only two course groups in one institutional setting and was concentrated mainly in Computer Engineering, which limits the generalizability of the findings to other disciplines, institutions, and cultural contexts. Second, depression and anxiety were measured using self-report instruments rather than clinical diagnosis, so the study identifies associations with reported symptom levels rather than clinically confirmed mental-health conditions. Third, the design was observational and essentially cross-sectional, which means that it cannot determine whether the observed behavioral patterns preceded psychological distress, resulted from it, or co-evolved with other unmeasured factors. In addition, the analysis examined a large number of extracted features, which raises the possibility that some associations may reflect chance findings even though several patterns were theoretically coherent and replicated across analyses.

A further limitation is that LMS behavior is always shaped by context. Course design, deadline structure, instructor practices, assessment type, workload intensity, and students' broader life circumstances can all influence when and how learners engage with the system. As a result, the same behavioral pattern may not carry the same meaning across all educational settings, and no single feature should be treated as a universal marker of depression or anxiety. This is especially important in mental-health-sensitive analytics, where false interpretation may create ethical risks as well as analytical error. Furthermore, the sample was predominantly male (80.4%), which may limit the generalizability of the findings to more gender-balanced student populations, particularly given established sex differences in the prevalence and behavioral expression of depression and anxiety.

Despite these limitations, the present study contributes several insights that justify its value. The observation that the same LMS behavioral feature, such as night-time log activity or session duration associates with depression in opposite directions across two course groups is not simply a replication failure. It constitutes direct empirical evidence that the behavioral expression of psychological distress in digital learning environments is moderated by course-level contextual factors, including academic year, workload structure, and assessment design. Prior single-cohort studies cannot detect this phenomenon by design, since they lack a within-institution comparative frame. By documenting both consistent and reversing patterns in parallel, this study provides a more honest and more actionable picture of the field: some LMS-derived indicators may serve as stable correlates of mental health across settings, while others require local calibration before any monitoring application is considered.

Future research should therefore test these findings in larger, more diverse, and more longitudinal datasets. Replication across disciplines, institutions, and cultural settings would help clarify which indicators are robust and which are context-dependent. Longitudinal studies would also be especially valuable because they could examine whether changes in temporal, session-based, and deadline-related behavior emerge before changes in depression and anxiety scores, rather than merely co-occurring with them. In addition, future work could explore multimodal approaches that combine LMS traces with other forms of data, as well as predictive models that evaluate whether behavioral indicators improve early detection beyond standard academic metrics. Equally important, future studies should involve students, instructors, and mental-health professionals in co-design processes so that any monitoring or intervention framework remains acceptable, transparent, and genuinely supportive. Future work should investigate comorbid depression and anxiety as a distinct analytical category. In the present study, a number of students met screening thresholds for both conditions simultaneously, yet were analyzed within separate group comparisons. Examining comorbid students as a dedicated group may reveal amplified or qualitatively distinct behavioral signatures in LMS data, and could inform the design of monitoring tools capable of distinguishing between pure and mixed mood presentations.

## VI. Conclusion

This study examined whether behavioral and performance-related indicators extracted from a Moodle-based electronic learning system were associated with university students' levels of depression and anxiety. The findings showed that several LMS-derived indicators were meaningfully related to these mental-health measures, especially temporal patterns of engagement, weekend activity, session duration, and deadline-related behavior. In particular, higher nighttime activity was associated with higher depression, greater weekend activity was linked to higher anxiety, and longer learning-session duration was related to higher depression.

The study contributes to the literature by extending learning analytics beyond traditional prediction of grades or dropout risk and toward the analysis of student well-being in digital learning environments. It also shows the value of combining behavioral, temporal, and performance-related LMS features with validated self-report measures of depression and anxiety within a single analytical framework. More broadly, the results suggest that the most informative signals of psychological strain may lie not simply in whether students are active in the system, but in how, when, and under what temporal conditions they engage with learning activities.

At the same time, these findings should be interpreted carefully. LMS-based behavioral indicators are not clinical evidence and should not be treated as diagnostic tools, but they may serve as useful contextual signals for early awareness and more supportive educational responses. For this reason, the main value of the present study lies in showing that routinely collected e-learning data can help reveal meaningful patterns related to depression and anxiety while preserving the essential distinction between behavioral indication and mental-health diagnosis.

## Conflict of Interest Statement

The authors declare that they have no relevant financial or non-financial competing interests to report. No employment, funding, personal fees, patents, stock ownership, collaborations, or any professional, ideological, or legal relationships exist that could be perceived as influencing the research presented in this article.

## Funding Declaration

No funding, grants, sponsorship, or financial support was received for conducting this research or for preparing this manuscript.

## APPENDIX A: Summary of Statistically Supported Behavioral and Performance Indicators

This appendix consolidates all behavioral and performance indicators for which a statistically significant group difference, a meaningful Spearman correlation with BDI/BAI scores or depression/anxiety status, or a notable effect size was observed. Table I reports, for each entry, the mean values for non-affected and affected students, the percentage difference, the t-test p-value, Cohen's d, and Spearman's ρ with BDI/BAI scores or binary status. Entries marked "n.s." denote p ≥ 0.05 but are retained for noteworthy effect sizes. Full feature extraction details are provided in Section III.

TABLE 1: SUMMARY OF BEHAVIORAL AND PERFORMANCE INDICATORS: GROUP DIFFERENCES, EFFECT SIZES, AND CORRELATIONS

| | # | Depression / Anxiety | Feature / Indicator | Normal Students (Mean) | Affected Students (Mean) | % Diff | p-value | Cohen's d | Spearman Correlation |
|---|---|---|---|---|---|---|---|---|---|
| Overall LMS Activity | 1 | Depression | Page Refresh Count | 55.43 | 41.83 | -24.5% | 0.0320 | -0.584 | ρ=-0.288, p=0.0256 |
| | 2 | Depression | Page Refresh Count Ratio | 10.88 | 9.03 | -17.0% | 0.0153 | -0.664 | ρ=-0.268, p=0.0383 |
| Temporal Engagement | 3 | Anxiety | Afternoon Log Count | 143.87 | 191.28 | +32.9% | 0.0084 | +0.705 | ρ=0.343, p=0.0073 |
| | 4 | Anxiety | Ratio of Afternoon Logs | 32.17 | 40.63 | +26.3% | 0.0128 | +0.663 | ρ=0.321, p=0.0125; BAI ρ=0.254, p=0.0499 |
| | 5 | Depression | Ratio of Afternoon Logs | 32.47 | 42.37 | +30.5% | 0.0043 | +0.789 | ρ=0.371, p=0.0035; BDI ρ=0.394, p=0.0018 |
| | 6 | Depression | Evening Log Count | 201.32 | 147.17 | -26.9% | 0.0315 | -0.585 | ρ = -0.261, p=0.0437 |
| | 7 | Depression | Night Log Count | 26.23 | 59.73 | +127.7 % | 0.0028 | +1.075 | BDI ρ=0.424, p=0.0090; status ρ=0.410, p=0.0117 |
| | 8 | Depression | Ratio of Night Logs | 7.32 | 16.80 | +129.5 % | 0.0167 | +0.842 | BDI ρ=0.409, p=0.0119; status ρ=0.380, p=0.0205 |
| | 9 | Depression | Weekend Log Count | 136.91 | 190.13 | +38.8% | 0.0245 | +0.787 | status ρ=+0.302, p=0.0696; BDI ρ=+0.322, p: n.s |
| | 10 | Depression | Weekend Log Ratio | 0.395 | 0.481 | +21.8% | 0.0283 | +0.766 | BDI ρ=0.371, p=0.0239; status ρ=0.343, p=0.0375 |
| | 11 | Anxiety | Weekend Log Ratio | 0.401 | 0.491 | +22.4% | 0.0292 | +0.798 | BAI ρ=0.501, p=0.0016; status ρ=0.363, p=0.0273 |
| Session-Based Patterns | 12 | Depression | Max Session Duration | 40,979 | 67,099 | +63.7% | 0.0029 | +1.072 | BDI ρ=0.498, p=0.0017; status ρ=0.495, p=0.0018 |
| | 13 | Depression | Total Session Duration | 137,559 | 227,377 | +65.2% | 0.0082 | +0.938 | status ρ=+0.314, p=0.0580; BDI ρ=+0.291, p=0.0806 |
| | 14 | Depression | Avg Session Duration | 5,027.64 | 7,461.87 | +48.4% | 0.0099 | +0.914 | status ρ=0.392, p=0.0165 |
| | 15 | Anxiety | Largest Inactive Period | 12.56 | 18.50 | +47.2% | 0.0082 | +0.984 | status ρ=+0.407, p=0.0124; BAI ρ=+0.203, p=0.2276 |
| | 16 | Anxiety | Max Session Duration | 45,209.04 | 64,817.08 | +43.3% | 0.0391 | +0.753 | status ρ=+0.314, p: n.s; |
| | | | | | | | | | BAI $\rho$=+0.319, p: n.s |
| | 17 | Anxiety | Avg Session Duration | 5,529.07 | 6,949.72 | +25.6% | 0.0293 | +0.577 | status $\rho$=+0.253, p: n.s;<br>BAI $\rho$=+0.170, p=0.1943 |
| **Deadline-Related Behavior** | 18 | Depression | HW1 Submission Lead Time (min) | 3,272 | -3,903 | -218% | 0.0031 | -0.818 | BDI $\rho$=-0.284, p=0.0280 |
| | 19 | Depression | HW2 Submission Lead Time (min) | 4,603 | 790 | -82.8% | 0.0378 | -0.565 | status $\rho$=-0.254, p=0.0498 |
| | 20 | Depression | HW1 Lead Time (log-transformed) | 2.87 | 1.55 | -45.8% | 0.0474 | -0.546 | BDI $\rho$=-0.274, p=0.0355 |
| | 21 | Anxiety | HW Lead Time (log-transformed) | 2.556 | 2.733 | +6.91% | n.s | +0.599 | status $\rho$=0.390, p=0.0172 |
| **Academic Performance / LMS Navigation** | 22 | Anxiety | Quiz Attempt Report Page Visits | 3.10 | 2.59 | -16.4% | 0.0146 | -0.650 | BAI $\rho$=-0.458, $p < 0.001$;<br>status $\rho$=-0.377, p=0.0030 |
| | 23 | Depression | HW Grade Announcement Visits | 9.05 | 7.96 | -12.1% | n.s | -0.157 | BDI $\rho$=-0.293, p=0.0231;<br>status $\rho$=-0.333, p=0.0094 |
| | 24 | Depression | Final Project Page Visits | 25.05 | 18.26 | -27.1% | 0.0142 | -0.671 | status $\rho$=-0.334, p=0.0091 |
| | 25 | Depression | Exam-related Activity Ratio | 0.333 | 0.267 | -19.8% | 0.0119 | -0.888 | BDI $\rho$=-0.391, p=0.0166;<br>status $\rho$=-0.361, p=0.0282 |
| | 26 | Depression | Number of HW Solution Page Visited | 2.59 | 4.47 | +72.4% | 0.0153 | +0.853 | BDI $\rho$=0.390, p=0.0169;<br>status $\rho$=0.343, p=0.0375 |
| | 27 | Depression | HW Grade Announcement Visits | 13.23 | 17.60 | +33.0% | n.s | +0.384 | BDI $\rho$=0.357, p=0.0299;<br>status $\rho$=0.342, p=0.0385 |
| | 28 | Depression | Final Grade Announcement Visits | 5.86 | 12.60 | +114.8 % | 0.0437 | +0.701 | BDI $\rho$=0.333, p=0.0442 |
| | 29 | Anxiety | Project Score | 81.44 | 98.42 | +20.8% | n.s | +0.622 | status $\rho$=+0.078, p: n.s;<br>BAI $\rho$=+0.227, p=0.177 |
| | 30 | Depression | Project Score | 80.05 | 97.07 | +21.2% | n.s | +0.627 | status $\rho$=+0.108, p: n.s;<br>BDI $\rho$=+0.257, p=0.125 |

- "status" in Spearman = correlation with the binary is_d / is_a grouping variable
- "BDI / BAI" in Spearman = correlation with the continuous inventory score
- "n.s." = not statistically significant